\documentclass[prl,twocolumn,longbibliography]{revtex4}

\usepackage[utf8]{inputenc}
\usepackage{bm}
\usepackage{mathtools}
\usepackage{bbold}
\usepackage{hyperref}
\usepackage[normalem]{ulem}
\usepackage{color}
\usepackage[dvipsnames]{xcolor}

\begin{document}

\title{Scalable spin squeezing from spontaneous breaking of a continuous symmetry}
\author{Tommaso Comparin$^1$, Fabio Mezzacapo$^1$, Martin Robert-de-Saint-Vincent$^2$,
Tommaso Roscilde$^1$}
\affiliation{$^1$ Univ Lyon, Ens de Lyon, CNRS, Laboratoire de Physique, F-69342 Lyon, France}
\affiliation{$^2$ Laboratoire de Physique des Lasers, Universit\'e Sorbonne Paris Nord F-93430 Villetaneuse, France and LPL CNRS, UMR 7538, F-93430 Villetaneuse, France}


\begin{abstract}
Spontaneous symmetry breaking (SSB) is a property of Hamiltonian equilibrium states which, in the thermodynamic limit, retain a finite average value of an order parameter even after a field coupled to it is adiabatically turned off. In the case of quantum spin models with continuous symmetry, we show that this adiabatic process is also accompanied by the suppression of the fluctuations of the symmetry generator -- namely, the collective spin component along an axis of symmetry. In systems of  $S=1/2$ spins or qubits, the combination of the suppression of fluctuations along one direction and of the persistence of transverse magnetization leads to spin squeezing -- a much sought-after property of quantum states, both for the purpose of entanglement detection as well as for metrological uses. Focusing on the case of XXZ models spontaneously breaking a U(1) (or even SU(2)) symmetry,  we show that the adiabatically prepared states have nearly minimal spin uncertainty; that the minimum phase uncertainty that one can achieve with these states scales as $N^{-3/4}$ with the number of spins $N$; and that this scaling is attained after an adiabatic preparation time scaling linearly with $N$.  Our findings open the door to the adiabatic preparation of strongly spin-squeezed states in a large variety of quantum many-body devices including \emph{e.g.} optical lattice clocks.  
\end{abstract}
\maketitle

\emph{Introduction.} Many-body entanglement \cite{Horodecki2009} is at the heart of the fundamental complexity of quantum states \cite{Popescuetal2006, Preskill2012Arxiv}, and it is the basis mechanism by which a quantum many-body system relaxes to a stationary state after having been driven away from equilibrium \cite{Kaufmanetal2016}. In this respect, the generation of many-body entangled quantum states is one of the main goals of a new generation of quantum devices, going from quantum simulators \cite{Georgescuetal2014,Altmanetal2021} to quantum computers \cite{nielsen_chuang_book}, whose common trait is the ability to perform coherent unitary evolutions of quantum many-body systems.  Yet, certifying (let alone putting to use) many-body entanglement is a task which is restricted to a small class of quantum many-body states, most prominently those whose entanglement can be detected via the measurement of the lowest moments of the quantum-noise distribution \cite{GuehneT2009}. In the context of ensembles of $N$ qubits ($S=1/2$ spins), characterized by the collective-spin operator ${\bm J} = \sum_{i=1}^N {\bm S}_i$ (where the ${\bm S}_i$'s are $S=1/2$ quantum spin operators), one of the best example of such states is offered by \emph{spin-squeezed} ones, whose entanglement is detected via the spin-squeezing parameter \cite{Wineland1994PRA}
\begin{equation}
\xi_R^2 = \frac{ N \min_{\perp} {\rm Var}(J^\perp)}{\langle {\bm J} \rangle^2},
\end{equation}
where $\min_{\perp}$ expresses the minimization of the variance on the collective-spin components perpendicular to the average spin orientation $\langle {\bm J} \rangle$. A state with $\xi_R^2 <1/k$ $(k\geq 1)$ is entangled, with an entanglement depth (least number of entangled spins) of $k+1$ \cite{Sorensen2001, Pezze2018RMP}. The $\xi_R^2$ parameter is also a fundamental figure of merit of the sensitivity of the state to rotations, expressing the reduction in phase-estimation error for a Ramsey interferometric protocol with respect to a factorized state \cite{Wineland1994PRA}; and it offers the possibility to improve the efficiency of quantum devices such as atomic clocks \cite{LouchetChauvetetal2010,Lerouxetal2010b,PedrozoPenafieletal2020,Schulteetal2020} or quantum sensors \cite{Muesseletal2014,Salvietal2018,Shankaretal2019} by using entanglement as a resource.

For the reasons listed above, devising many-body mechanisms that lead to the controlled preparation of spin-squeezed states \cite{Ma2011PR} is a very significant endeavor -- and particularly so when the squeezing parameter can be parametrically reduced by increasing the number of resources. This situation leads to \emph{scalable squeezing} -- generically $\xi^2_R \sim N^{-\alpha}$ ($\alpha>0$) -- which allows one to surpass the standard quantum limit for the scaling of the phase-estimation error with the number of qubits \cite{Pezze2018RMP}.  The two main mechanisms that have been identified and implemented so far in this direction are the preparation of spin-squeezed states via unitary evolutions with long-range interactions \cite{Kitagawa1993PRA,Esteve2008,Riedel2010,Bohnet2016,Perlin2020PRL} and via the non-demolition measurement of a spin component \cite{Lerouxetal2010,Hosten2016}. Yet a third way to spin squeezing is offered by \emph{adiabatic preparation}, which relies upon the identification of spin squeezing in the ground state of quantum spin Hamiltonians implemented \emph{e.g.} by quantum simulation platforms. A clear example in this direction is offered by Ising quantum critical points \cite{Vidaletal2004}, which exhibit scalable squeezing at and above the upper critical dimension ($d\geq 3$ for short-range interactions) \cite{Frerot2018PRL_B}. 

\begin{figure}[ht!]
\begin{center}
\includegraphics[width=0.9\columnwidth]{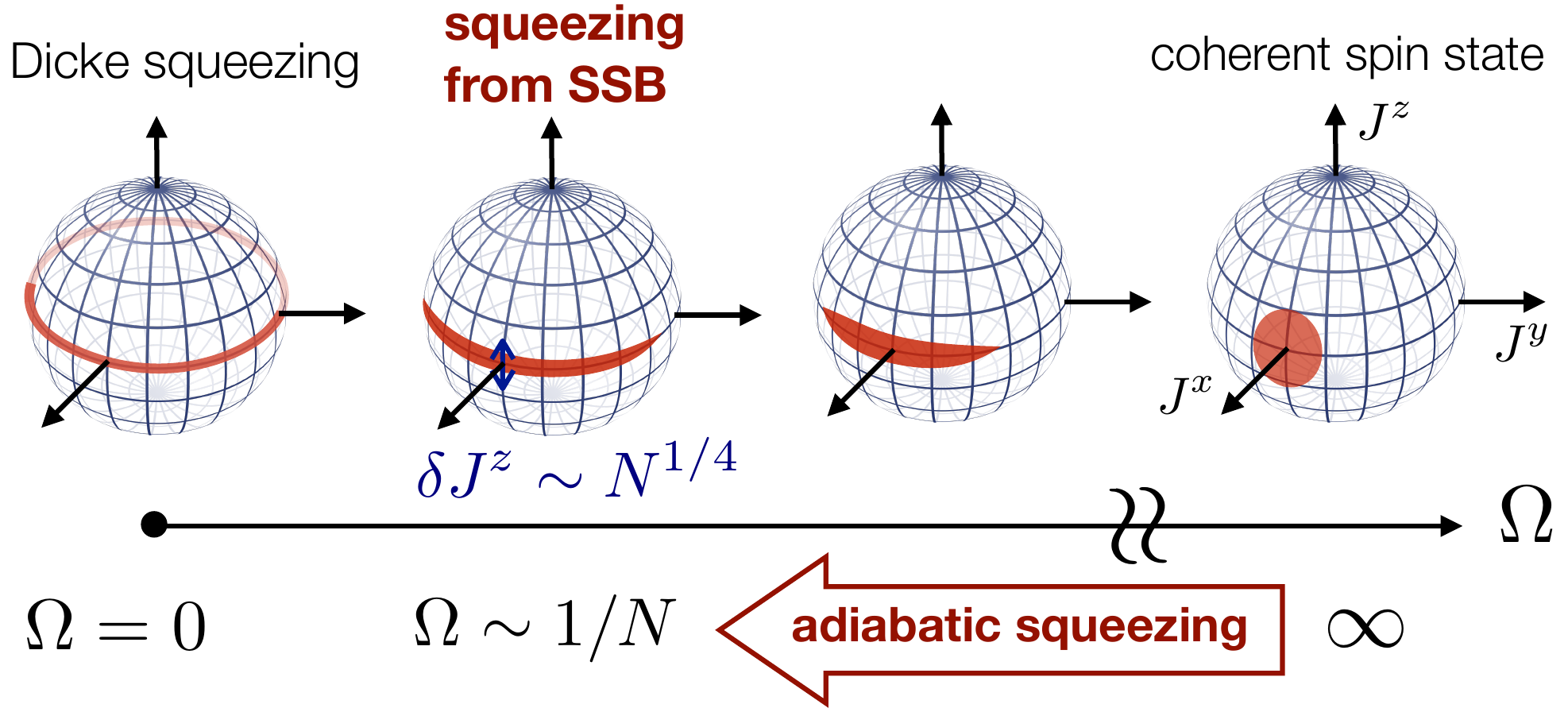}
\caption{\emph{Adiabatic squeezing from spontaneous symmetry breaking in the XXZ model.}  Starting from a coherent spin state at $\Omega = \infty$, an adiabatic reduction of the field $\Omega$ coupling to the order parameter leads to the appearance of scalable spin squeezing when $\Omega \sim 1/N$, due to the scaling of the uncertainty on the  $J^z$ component, $\delta J^z = \sqrt{{\rm Var}(J^z)} \sim N^{1/4}$; and to the absence of scaling of the order parameter $m=\langle J^x \rangle/N$, as a consequence of spontaneous symmetry breaking (SSB). The red areas depict the uncertainty regions of the collective spin on a sphere of radius $\sqrt{\langle \bm J^2 \rangle} \sim N$. As a consequence the angular aperture of the uncertainty region along the $z$ axis is $\delta \phi \approx \delta J^z /\sqrt{\langle \bm J^2 \rangle} \sim N^{-3/4}$, defining the sensitivity of the state to rotations around the $y$ axis.}
\label{f.sketch}
\end{center}
\end{figure}

In this work we unveil another fundamental link between many-body physics of spin models and the generation of spin squeezing, namely the appearance of squeezing in the presence of \emph{spontaneous breaking of a continuous spin symmetry} in the ground state. Without loss of generality, in the following we will be concerned with symmetry under U(1) rotations $U_z(\phi) = e^{-i\phi J^z}$ generated by the collective spin component $J^z$ -- this property is also present for SU(2)-symmetric Hamiltonians. On a finite-size system and in the absence of any symmetry-breaking field, the ground state of a U(1) symmetric Hamiltonian  has ${\rm Var}(J^z) = 0$, namely it exhibits so-called Dicke squeezing \cite{Pezze2018RMP} {(Fig~\ref{f.sketch}(a))}, lacking nonetheless a finite net magnetization $\langle {\bm J} \rangle =0$. At the same time, 
the low-lying energy spectrum of that same Hamiltonian exhibits a so-called Anderson tower of states (ToS), whose energy collapses as $1/N$ onto that of the ground state \cite{Anderson1997,Tasaki2018JSP,Beekman2019SPPLN,Comparinetal2021}. Hence a field $\Omega \sim 1/N$ coupling to the order parameter in the $x-y$ plane, e.g.   $- \Omega J^x$ (without loss of generality), is sufficient to mix the ToS into a state exhibiting a net polarization $m = \langle J^x \rangle/N \neq 0$ {(Fig~\ref{f.sketch}(b))}. The hallmark of spontaneous symmetry breaking (SSB) is then the persistence of a finite order parameter $m$ in the limit $N\to\infty$, in which the field is also parametrically set to zero. Here we investigate paradigmatic XXZ models with nearest-neighbor interactions using finite-temperature and variational quantum Monte Carlo simulations, as well as of spin-wave theory; and, in the presence of SSB in the ground state, we show that the state polarized by the minimal field $\Omega \sim 1/N$ away from Dicke squeezing retains a strong asymmetry in the fluctuations of the collective spin components, exhibiting scalable (Wineland) spin squeezing with $\xi_R^2 \sim N^{-1/2}$. Such a state is shown to have minimal spin uncertainty, namely squeezing is its optimal metrological resource; and it can be prepared adiabatically, starting from a coherent spin state stabilized at $\Omega \to \infty$ {(Fig~\ref{f.sketch}(d))}, and ramping down $\Omega$ to a value $\sim 1/N$ in a time scaling linearly with system size, $\tau \sim N$. This finding opens the possibility to squeeze the collective spin of quantum simulators of U(1)- (or SU(2)-)symmetric qubit Hamiltonians, with potential applications to quantum sensors \cite{Muesseletal2014,Faconetal2016} and atomic clocks \cite{Takamotoetal2005, Campbelletal2017}. 


\begin{figure*}[ht!]
\begin{center}
\includegraphics[width=0.9\textwidth]{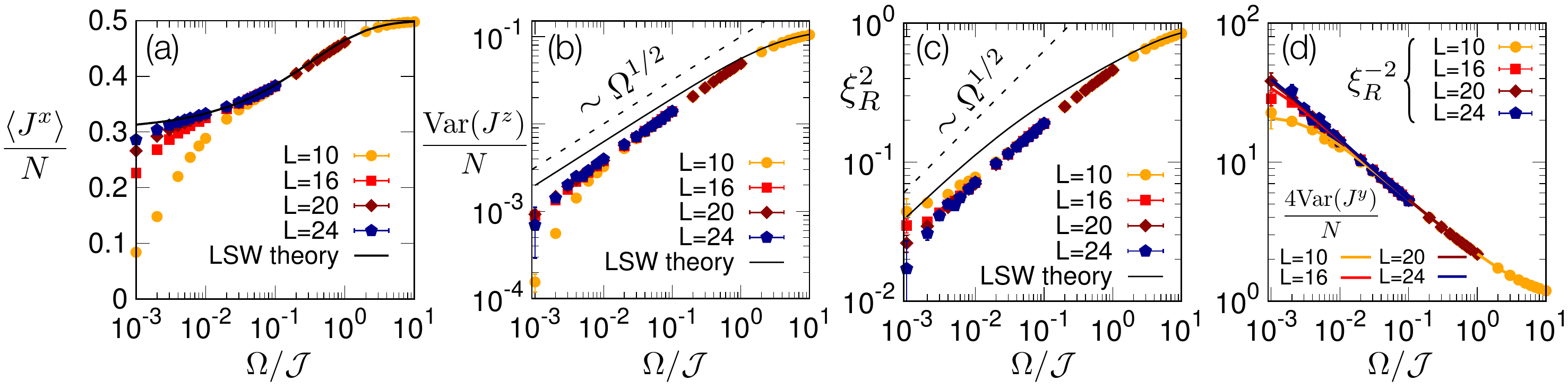}
\caption{\emph{Adiabatic squeezing from SSB in the 2$d$ Heisenberg model.}  (a) Field-induced magnetization $\langle J^x\rangle/N$ for various lattice sizes $N= L^2$; (b) Variance of the collective spin component $J^z$; (c) resulting spin-squeezing parameter $\xi_R^2$; (d) comparison between $\xi_R^{-2}$ and $4 {\rm Var}(J^y)/N$. In all panels the solid black line indicates the prediction of linear spin-wave (LSW) theory, and the dotted line in panels (b)-(c) shows the $\Omega^{1/2}$ scaling as a reference (multiplied by an arbitrary prefactor). Here and in the following figures, the error bars denote one standard deviation of the statistical fluctuations in QMC.}
\label{f.2dXXZ}
\end{center}
\end{figure*}

\emph{Model and methods.} We focus our attention on the $S=1/2$ XXZ Hamiltonian:
\begin{equation}
{\cal H} = - \frac{1}{2} \sum_{ij} {\cal J}_{ij} (S^x_i S^x_j + S^y_i S^y_j - \Delta S^z_i S^z_j ) - \Omega \sum_i \epsilon_i S_i^x   
\end{equation}
where $i,j$ are lattice sites on a $d$-dimensional hypercubic lattice of size $N = L^d$ with periodic boundary conditions. In the following we shall specialize our attention to nearest-neighbor (n.n.) interactions -- ${\cal J}_{ij} = {\cal J}$ if $i$ n.n. $j$ (see Supplemental Material - SM -  for an extension to longer-range interactions \cite{SM}). Moreover we choose ${\cal J}>0$ and  $-1 < \Delta \leq 1$, defining an XXZ model with ferromagnetic interactions in the plane and ferromagnetic or antiferromagnetic along the symmetry axis. Such a situation is realized e.g. in bosonic Mott insulators \cite{Duanetal2003,Jepsen2020,Sunetal2021}. Under these assumptions, the above model is known to break a continuous symmetry in the ground state in $d\geq 2$; 
and the field coupling to the order parameter is uniform, namely $\epsilon_i = 1 ~\forall i$. But most of our results are completely general, and apply to any model featuring SSB of a continuous symmetry -- provided that the order parameter \emph{does not} commute with the Hamiltonian (otherwise the ground state is simply a coherent spin state for all $\Omega$, e.g. $\otimes_{i=1}^N |\rightarrow_x\rangle_i$ with $|\rightarrow_x\rangle_i$ the eigenstate of $S_i^x$ with eigenvalue $+1/2$). In the case ${\cal J}<0$ (antiferromagnetic interactions in the $xy$ plane) -- as realized in fermionic Mott insulators \cite{Duanetal2003, Greifetal2013, Mazurenko2017} -- the order parameter is the staggered magnetization, \emph{e.g.}  $m = \langle J^x_{\rm st}\rangle /N = \langle \sum_i \epsilon_i S_i^x \rangle/N$; the field coupling to the order parameter must therefore be staggered, $\epsilon_i = (-1)^i$; and the relevant rotations are generated by  $J^y_{\rm st} = \sum_i \epsilon_i S_i^y$. Yet, for n.n. interactions on a hypercubic lattice, the physics is equivalent to that of the bosonic insulators, as the two models are connected by a canonical transformation (rotation of $\pi$ around the $z$ axis for one of the two sublattices).

We have studied the ground-state physics of the XXZ Hamiltonian in $d=1, 2$ and 3 making use of numerically exact quantum Monte Carlo (QMC) simulations, based on the Stochastic Series Expansion method \cite{Syljuasen2002PRE}; as well as of spin-wave theory, valid in the presence of spontaneous symmetry breaking ($d=2,3$). Moreover we have investigated the (quasi-)adiabatic dynamics of preparation of the ground state starting from a large $\Omega$ by making use of time-dependent variational Monte Carlo (tVMC), based on pair-product (or spin-Jastrow) wavefunctions \cite{Thibaut2019PRB,Comparinetal2021} as well as of time-dependent spin-wave theory (see SM for an extended discussion of the methods \cite{SM}).

\emph{Adiabatic squeezing from SSB.} In the following we shall only show results for the case $\Delta = 1$ (Heisenberg model) -- analogous results for different $\Delta$ values are presented in the SM. Fig.~\ref{f.2dXXZ} shows our QMC results for the 2$d$ Heisenberg model calculated for different sizes $N=L^2$ at a temperature  $T = {\cal J}/N$ chosen so as to effectively remove thermal effects at the energy scale of an applied field $\Omega \sim {\cal J}/N$. As we shall see, this choice is rather conservative, because the field opens in fact a gap in the spectrum scaling as $({\cal J} \Omega)^{1/2}$. The QMC results are compared to linear spin-wave (LSW) theory -- see SM for the details of the theory --  in the thermodynamic limit, which is expected to be very accurate at large $\Omega$, and to quantitatively capture some selected features in the limit $\Omega \to 0$ \cite{Manousakis1991}. The uniform magnetization $\langle J^x \rangle/N$, shown in Fig.~\ref{f.2dXXZ}(a), is indeed correctly predicted by LSW theory: the finite-size QMC data show that the LSW prediction is reproduced down to a field scaling $\sim 1/N$, below which the finite-size gap between the states of the Anderson tower overcomes the field {and the uniform magnetization is strongly suppressed}. Hence the SSB scenario, namely the persistence of a finite magnetization down to $\Omega =0$ when $N \to \infty$, is clearly shown. Concomitantly, the suppression of the symmetry breaking field $\Omega$ leads to a strong suppression of the fluctuations of the U(1) symmetry generator $J^z$; LSW theory predicts that ${\rm Var}(J^z) \sim \Omega^{1/2}$ when $\Omega \to 0$, a prediction which appears to be consistent with our finite-size QMC results for \emph{e.g.} the 2$d$ XX model (see Ref.~\cite{SM}) for fields down to $\Omega \approx {\cal J}/N$; while the 2$d$ Heisenberg model shows significant beyond-LSW corrections, which interestingly appear to lead to a further reduction of the variance, namely to stronger squeezing \cite{SM}. The combination of these two results implies naturally that the $\xi_R^2$ parameter is smaller than unity for all values of $\Omega$ -- in agreement with a recent theorem predicting ground-state squeezing in this model as soon as $\Omega < \infty$ \cite{Roscildeetal2021}; and which scales as $\Omega^{1/2}$ (actually faster for the 2$d$ Heisenberg model) down to fields $\sim {\cal J}/{N}$, namely as $N^{-1/2}$ (or faster) for the lowest significant fields for each finite size $N$. The evidence of scalable spin squeezing resulting from SSB -- namely from the absence of scaling (or persistence) of $\langle J^x \rangle/N$ -- is the main result of our work. 

\begin{figure*}[ht!]
\begin{center}
\includegraphics[width=0.9\textwidth]{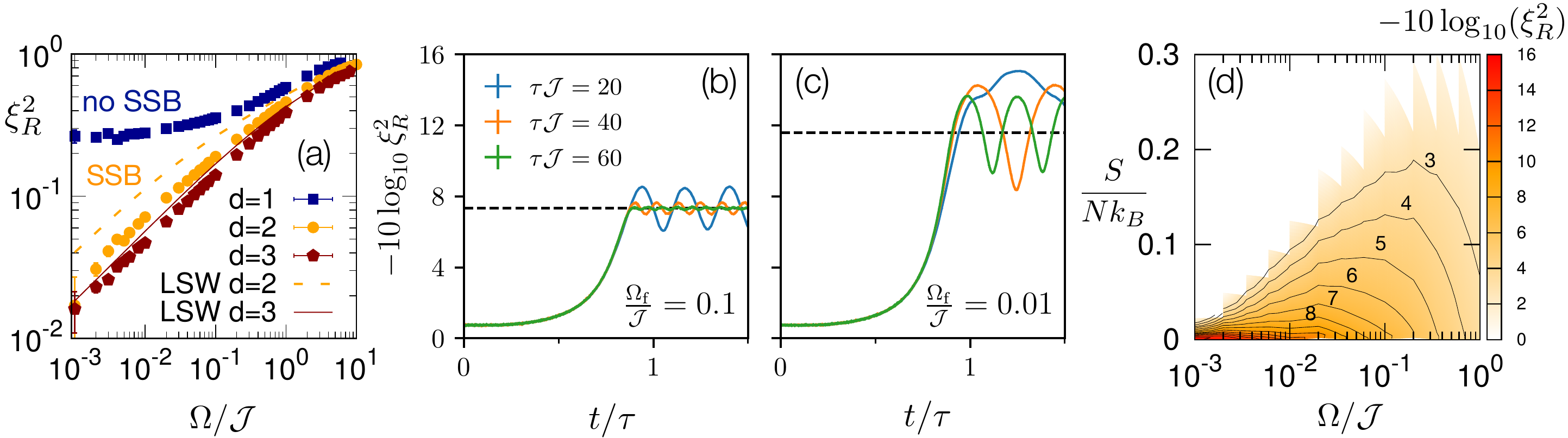}
\caption{\emph{Quasi-adiabatic preparation of the low-field state.} (a) Comparison between the field dependence of the squeezing parameter $\xi_R^2$ for the ground state of the Heisenberg model in $d=$1, 2, and 3. For each value of $\Omega$, we use a system size $N$ such that $\Omega \geq {\cal J}/N$, at a temperature $T/{\cal J} = 1/N$ removing thermal effects. The dashed and solid lines show the prediction of LSW theory. (b-c) The two panels show tVMC results for the evolution of the spin-squeezing parameter in the 2$d$ Heisenberg model ($L=12$) along two field ramps starting from $\Omega_i/{\cal J} = 10$ and ending at  (b)  $\Omega_{\rm f}/{\cal J} = 10^{-1}$ and (c) $10^{-2}$ -- see text for the ramp protocol. Each panel shows three different ramps for different ramp durations $\tau$. The dashed lines show the ground-state spin squeezing parameters -- obtained by variational minimization of the energy with the spin-Jastrow Ansatz. (d) Spin squeezing parameter vs. applied field and entropy per spin in the 2$d$ Heisenberg model, $L=24$.}
\label{f.ramp}
\end{center}
\end{figure*}

Another significant feature of the low-$\Omega$ state of XXZ models exhibiting SSB is that of being a state of \emph{minimal uncertainty} for the collective spin, namely the collective spin components saturate the Heisenberg-Robertson inequality, ${\rm Var}(J^y) {\rm Var}(J^z) \geq \langle J^x \rangle^2/4$. To discuss this aspect and its metrological implications, it is useful to introduce the quantum Fisher information (QFI) density \cite{Pezze2018RMP} for the $J^y$ component, defined as $f(J^y) = \frac{2}{N} \sum_{nm} \frac{(p_n-p_m)^2}{p_n+p_m} |\langle m |J^y|n\rangle|^2$, where $|n\rangle$ and $p_n$ are the eigenstates and corresponding eigenvalues of the density matrix $\rho$. When the state in question is rotated around the $y$ axis by the transformation $U_y(\phi) = e^{-i\phi J^y}$, the QFI density expresses the minimal uncertainty on the angle $\phi$, $\delta \phi \geq (f_Q N)^{-1/2}$, namely $f_Q\neq 1$ implies a deviation of this uncertainty with respect to the standard quantum limit. The latter property, combined with the fact that $4{\rm Var}(J^y)/N$ is an upper bound to the QFI density, leads to the inequality chain: 
\begin{equation}
\xi_R^{-2} = \frac{\langle J^x \rangle^2}{N {\rm Var}(J^z)}\leq f_Q(J^y) \leq \frac{4 {\rm Var}(J^y)}{N} ~.
\end{equation}
If a state has minimal uncertainty, namely ${\rm Var}(J^y) {\rm Var}(J^z) \simeq \langle J^x \rangle^2/4$, the above inequality chain collapses to an equality, namely $\xi_R^{-2} \approx f_Q(J^y) \approx 4 {\rm Var}(J^y)/N$. This collapse is clearly exhibited by our numerical data for all values of $\Omega>0$ and all systems sizes -- see Fig.~\ref{f.2dXXZ}(d).
 In particular, among all the macroscopic observables built as a sum of local observables, $J^y$ is arguably the one with the largest QFI density \cite{SM}, so that the estimation of the rotation angle $\phi$ is the optimal phase-estimation protocol for the low-$\Omega$ states. The fact that $\xi_R^{-2} \approx f_Q$ implies that the measurement of the rotation of the average collective spin (corresponding to Ramsey interferometry) is the \emph{optimal measurement} for this protocol, leading to a phase-estimation error $\delta \phi = \xi_R/\sqrt{N} \sim N^{-3/4}.$ 

Finally, we would like to stress that the above results are not at all limited to the 2$d$ Heisenberg model, but they are valid for all the $S=1/2$ XXZ models spontaneously breaking a U(1) (or SU(2)) symmetry in the thermodynamic limit (see \cite{SM} for further examples). Fig.~\ref{f.ramp}(a) shows the field dependence of the spin-squeezing parameters $\xi_R^2$ for the Heisenberg model in $d$=1, 2, and 3. We observe that the scaling of $\xi_R^2$ as $\Omega^{1/2}$ is clearly exhibited in $d=3$. On the other hand, for $d=1$ (Heisenberg chain) SSB is not realized because of the critical strength of quantum fluctuations \cite{Auerbach1994}: as a consequence, $\langle J^x \rangle$ vanishes when $\Omega \to 0$, leading to the breakdown of the mechanism that underpins scalable spin squeezing in higher dimensions. Fig.~\ref{f.ramp}(a) shows that $\langle J^x \rangle^2$, vanishing as $\Omega^{1/2}$, leads to a squeezing parameter $\xi_R^2$ that goes to a constant as $\Omega \to 0$. 
Similar results for the XX model ($\Delta=0$) are shown in the SM \cite{SM}.

\emph{Quasi-adiabatic ramps.} The preparation of the ground state at low fields requires the initialization of the system in a coherent spin-state aligned with the $\Omega$ field with $\Omega \gg {\cal J}$; and the subsequent gradual reduction of the field along an adiabatic down-ramp -- a protocol analog to that of adiabatic quantum computing \cite{AlbashL2018}. The adiabatic theorem mandates that the duration $\tau$ of an adiabatic ramp that prepares the system in the ground state at a final field $\Omega_{\rm f}$ should be $\tau {\cal J} \gtrsim (\Delta E_{\rm min}/{\cal J})^{-2}$, where $\Delta E_{\rm min} = \min_{\Omega\in [\Omega_{\rm f},\infty]} ( E_1(\Omega) - E_0(\Omega))$ is the minimal gap between the $\Omega$-dependent ground-state energy ($E_0$) and the energy of the first excited state ($E_1$) over the field range of the ramp. This gap can be calculated by LSW theory \cite{SM} --  in good agreement with exact diagonalization on small system sizes \cite{SM} -- and for the Heisenberg model ($\Delta = 1$) and $\Omega_{\rm f} \ll {\cal J}$ it is shown to be $\Delta E_{\rm min}/{\cal J} \approx (z \Omega_{\rm f}/{\cal J})^{1/2}$ where $z=2d$ is the coordination number. This result implies that the adiabatic preparation of the ground state for the minimal field $\Omega_{\rm f} \sim 1/N$ at size $N$ takes a time $\tau/{\cal J} \gtrsim (z \Omega_{\rm f}/J)^{-1}  \sim N$. 

We complement the above general prediction from LSW theory with realistic calculations of quasi-adiabatic ramps based on tVMC -- which show remarkable agreement with independent calculations based on time-dependent LSW \cite{SM}, mutually corroborating their quantitative validity. We start the state evolution from the ground state at a large initial field value $\Omega_{\rm i} = 10 {\cal J}$ -- obtained by minimization of the variational energy of the spin-Jastrow Ansatz \cite{SM}; and then we ramp the field down to $\Omega_{\rm f}$ with the schedule $\Omega(t) = \Omega_{\rm i} + F(t/\tau) (\Omega_{\rm f}-\Omega_{\rm i})$, where 
$F(x) =  \frac{1}{2} e^{-1/x+2} ~\theta(1/2-x) + \left [ 1- \frac{1}{2} e^{-1/(1 - x)+2} \right ] \theta(x-1/2)$
for $t \in [0,\tau]$, while $\Omega(t) = \Omega_{\rm f}$ for $t>\tau$. The function $F(t)$ {(chosen heuristically)} has the property of having vanishing derivatives at all orders at the two extremes of the $[0,\tau]$ interval, so that it is continuous along with all of its derivatives when it is extended to $t<0$ and $t>\tau$ by constant functions. Fig.~\ref{f.ramp}(b-c) shows the tVMC results for evolution of the $\xi_R^2$ parameter in the 2$d$ Heisenberg model ($L=12$) with two different final fields ($\Omega_{\rm f}/{\cal J} = 10^{-1}$ and $10^{-2}$), and various ramp durations. Our main observation is that, even when the ramp fails to keep the system in its ground state down to $\Omega_{\rm f}$, the squeezing parameter exceeds the adiabatic value only for $t\lesssim \tau$, while it systematically evolves to lower values at immediately later times, and then oscillates around the adiabatic value. Therefore failure to follow a perfectly adiabatically ramp (which in Fig.~\ref{f.ramp} is observed for all considered ramp durations when $\Omega_f = 10^{-2} {\cal J}$) does not \emph{per se} imply a degradation of the amount of squeezing that can be produced in the system.  

A final comment concerns the possibly of imperfect preparation of the initial state of the quasi-adiabatic ramp: this would generically entail the presence of finite entropy in the initial state, persisting then in the evolved one. Fig.~\ref{f.ramp}(d) tests the robustness of squeezing to the presence of finite entropy in the case of the equilibrium state of the 2$d$ Heisenberg model. Not surprisingly, a finite entropy imposes a limit to the achievable squeezing; yet adiabatic spin squeezing can be obtained up to spin entropies $S/N \lesssim 0.3 k_B$.

\emph{Conclusions.}  In this work we have demonstrated a fundamental mechanism for the equilibrium preparation of many-qubit entangled states featuring scalable spin squeezing, based on the adiabatic preparation of low-field magnetized ground states for Hamiltonians breaking a continuous (U(1) or SU(2)) symmetry in the thermodynamic limit. At variance with the existing schemes for spin squeezing using collective-spin interactions \cite{Kitagawa1993PRA,Esteve2008,Riedel2010,Bohnet2016,Lerouxetal2010,Hosten2016}, here we offer a specific protocol for the production of scalable spin squeezing using \emph{short-range} qubit Hamiltonians with continuous symmetry, whose implementation is common to nearly all quantum simulation platforms. Our results are immediately relevant for Mott insulators of bosonic ultracold atoms in optical lattices, realizing the XXZ model with SU(2) symmetry or U(1) symmetry (easy-plane anisotropy)  -- see e.g. the two relevant cases of $^7$Li \cite{Jepsen2020} and of $^{87}$Rb \cite{Sunetal2021}); and to Mott insulators of fermionic atoms, realizing the Heisenberg antiferromagnet \cite{Greifetal2013,Mazurenko2017}. In the bosonic case the $\Omega$ field coupled to the order parameter is a uniform, coherent Rabi coupling between two internal states; while in the fermionic case the field coupling to the order parameter must be \emph{staggered}, and it can be potentially created by Stark shifting a sublattice of a square or cubic lattice by a superlattice, therefore creating a Rabi-frequency difference between the two sublattices. This scheme opens the possibility to squeeze the spin state of optical-lattice clocks in the Mott insulating regime (e.g. based on $^{87}$Sr \cite{Takamotoetal2005, Campbelletal2017} {in the fermionic case, and on $^{174}$Yb in the bosonic case \cite{Bouganneetal2017, Franchietal2017} -- see SM for further discussion}). Our protocol (with a uniform Rabi field $\Omega$) is also relevant for superconducting circuits realizing \emph{e.g.} the 2$d$ XX Hamiltonian \cite{Mi2021}; for Rydberg atoms with resonant interactions \cite{Browaeys2020NP}, realizing the dipolar XX model {$\Delta = 0$, ${\cal J}_{ij} \sim |r_i - r_j|^{\alpha}$} with $\alpha=3$; as well as for trapped ions, realizing the XX model with long-range interactions ($0< \alpha <3$) \cite{Brydges2019}. Our findings pave the way for the controlled adiabatic preparation of scalable spin-squeezed states, with the double bonus of a solid entanglement certification via the measurement of the collective spin; and of the possibility to accelerate the size scaling of phase-estimation error compared to separable states. 

\begin{acknowledgments} \emph{Acknowledgements.} We acknowledge useful discussions with M. Tarallo and G. Bertaina. We particularly thank B. Laburthe-Tolra and L. Vernac for their contributions to the conception of this project, and for their careful reading of our manuscript. This work is supported by ANR (``EELS" project) and by QuantERA (``MAQS" project). All numerical simulations have been performed on the PSMN cluster of the ENS of Lyon. 
\end{acknowledgments}

\newpage

\begin{center}
{\bf Supplemental Material } \\
{\bf \emph{Scalable spin squeezing from spontaneous breaking of a continuous symmetry}}
\end{center}

\section{Spin-wave theory for the XXZ model}
\label{s.LSW}

In this section we describe the well-known linear spin-wave (LSW) theory as applied to the XXZ model in an applied field. 
We introduce the linearized Holstein-Primakoff transformation 
\begin{eqnarray}
S^x_i & = & \frac{1}{2} - n_i  \\
S^y_i & = & \frac{1}{2} \left ( \sqrt{1-n_i} ~b_i + b_i^\dagger \sqrt{1-n_i} \right) \approx \frac{1}{2} \left ( b_i + b_i^\dagger \right ) \nonumber \\
S^z_i & = & \frac{1}{2i} \left ( \sqrt{1-n_i} ~b_i + b_i^\dagger \sqrt{1-n_i} \right) \approx \frac{1}{2i} \left ( b_i - b_i^\dagger \right ) \nonumber
\end{eqnarray}
in which $n_i = b_i^\dagger b_i$ and $b_i, b_i^\dagger$ are bosonic destruction and creation operators -- where the linearization assumes that $\langle n_i \rangle \ll 1$ for all quantum states of interest.    
Under this transformation the XXZ Hamiltonian takes the form of a quadratic bosonic Hamiltonian
\begin{equation}
{\cal H} \approx E_0 +  \frac{1}{2} \sum_{\bm k} \left [ 2 A_{\bm k} b_{\bm k}^\dagger b_{\bm k} + B_{\bm k} (b_{\bm k} b_{-\bm k} + {\rm h.c.}) \right ]
\label{e.LSW}
\end{equation}
where we have introduced the Fourier transformed bosonic operators
\begin{equation}
b_{\bm k} = \frac{1}{\sqrt{N}} \sum_i e^{-i \bm k \cdot \bm r_i} b_i~;
\end{equation}
the mean-field energy
\begin{equation}
E_0 = - \frac{\sum_{ij} {\cal J}_{ij}}{4} - \frac{N \Omega}{2};
\end{equation} 
and the coefficients 
\begin{eqnarray}
A_{\bm k} &=& \gamma_0 + (\Delta-1) \frac{\gamma_{\bm k}}{2} + \Omega \nonumber \\
B_{\bm k} &=& (\Delta + 1) \frac{\gamma_{\bm k}}{2} \nonumber \\
\gamma_{\bm k} & =  &\frac{1}{N} \sum_{ij} {\cal J}_{ij} e^{i {\bm k}\cdot ({\bm r}_i - {\bm r}_j)}~.
\end{eqnarray}

\begin{figure}[ht!]
\begin{center}
\includegraphics[width=0.9\columnwidth]{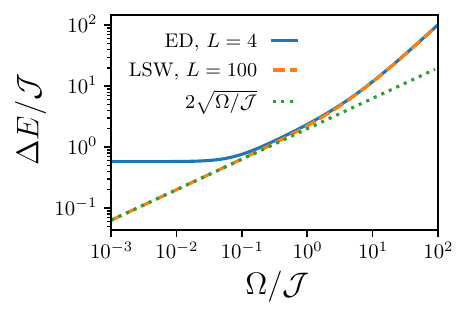}
\caption{Energy gap of the 2$d$ Heisenberg model as a function of the applied $\Omega$ field: the prediction of LSW theory for a large system ($L=100$) is compared to that of exact diagonalization (ED) for a small system ($L=4$), and to the expected low-field behavior $\Delta E \approx 2{\cal J} \sqrt{\Omega/{\cal J}}$, valid in the thermodynamic limit. }
\label{f.gap}
\end{center}
\end{figure}

The above Hamiltonian is diagonalized via the Bogoliubov transformation 
\begin{equation}
b_{\bm k} = u_{\bm k} \beta_{\bm k} - v_{\bm k} \beta^\dagger_{-\bm k}
\end{equation}
into 
\begin{equation}
{\cal H} = E_0 + \sum_{\bm k} \epsilon_{\bm k} \beta_{\bm k}^\dagger \beta_{\bm k} + {\rm const.} 
\end{equation}
where $\epsilon_{\bm k} = \sqrt{A_{\bm k}^2 - B_{\bm k}^2}$ and 
\begin{equation}
u_{\bm k} = \frac{1}{\sqrt{2}} \left ( \frac{A_{\bm k}}{\epsilon_{\bm k}} + 1 \right )^{1/2}
~~~ v_{\bm k} = {\rm sign}(B_{\bm k}) (1-u_{\bm k}^2)^{1/2} ~.
\end{equation}

The observables relevant for the calculation of the squeezing parameter are then
\begin{eqnarray}
\langle J^x \rangle & = & \frac{N}{2} - \frac{1}{2} \sum_{\bm k} v_{\bm k}^2 = N - \sum_{\bm k} \frac{A_{\bm k}}{2\epsilon_{\bm k}} \\
{\rm Var}(J^z) & = &  \frac{N}{4} \left (1 + 2 v_0^2 + u_0 v_0 \right ) = \frac{N}{4} \sqrt{\frac{ A_0 + B_0}{A_0 - B_0}}  \nonumber~.
\end{eqnarray}

The comparison between the predictions of LSW theory and QMC is already shown in the main text, as well as in the next section. 
Fig.~\ref{f.gap} shows the LSW prediction for the spectral gap $\epsilon_{\bm k = 0}$, compared to the exact diagonalization (ED) for a $N=16$ square lattice, in the case of the SU(2) model with $\Delta = 1$. The exact results were obtained via the QuSpin package \cite{Weinberg2017SP,Weinberg2019SP}. We observe that the gap predicted by LSW theory is very accurate at large fields; the deviation from the ED results at lower field is due to the saturation of the ED gap to a value $\sim {\cal O}(N^{-1})$ due to the finite size, while the LSW gap closes as $\Omega^{1/2}$ when $\Omega \to 0$.

\begin{figure*}[ht!]
\begin{center}
\includegraphics[width=0.8\textwidth]{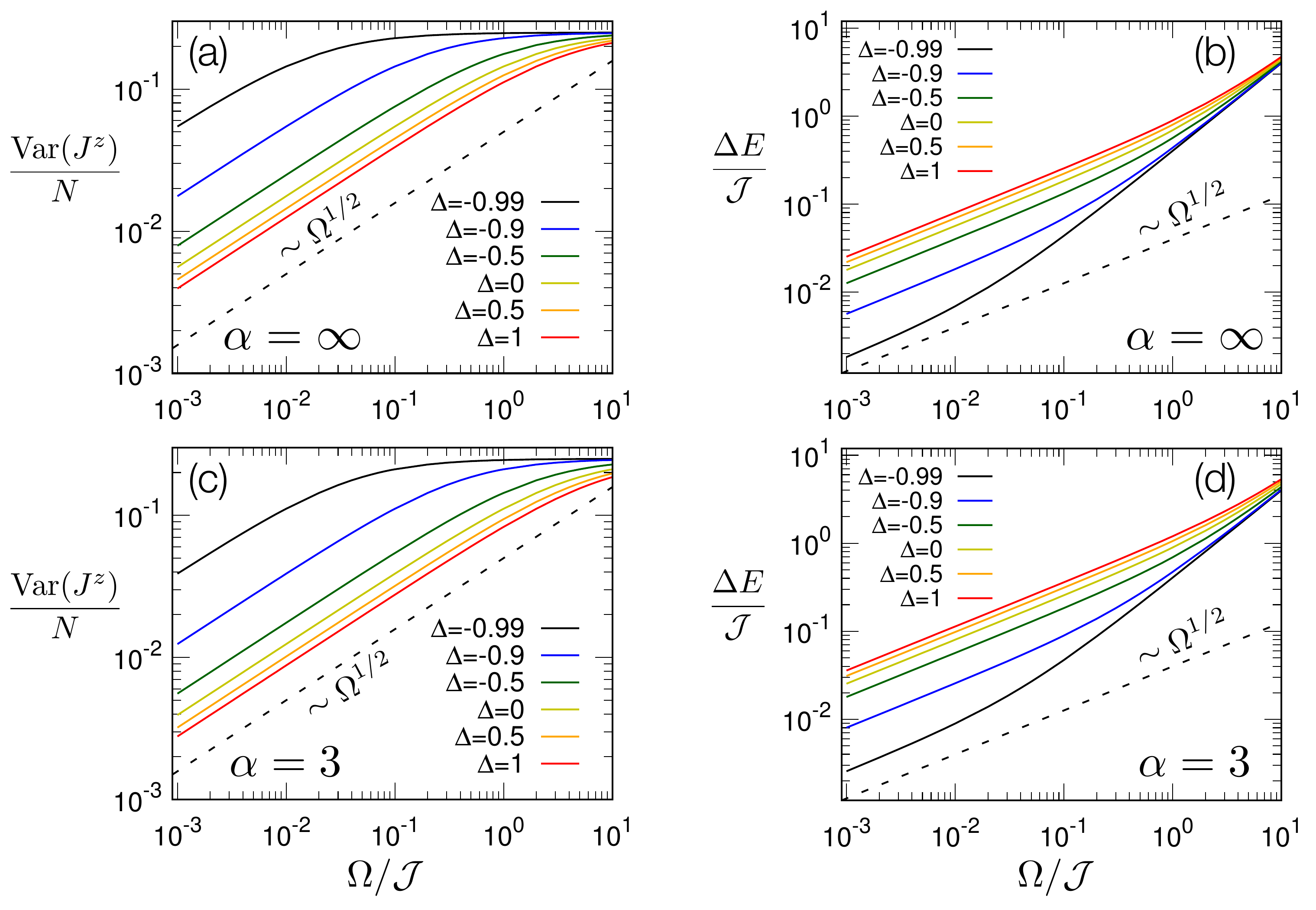}
\caption{LSW predictions for the 2d XXZ model with variable anisotropy and interaction range: (a) ${\rm Var}(J^z)/N$  and (b) excitation gap $\Delta E/{\cal J}$ for various anisotropies $\Delta$ in the 2d XXZ with nearest-neighbor interactions ($\alpha=\infty$); (c) ${\rm Var}(J^z)/N$  and (d) excitation gap $\Delta E/{\cal J}$ for the same model with dipolar interactions $\alpha=3$. In all the panels the dashed line indicates a $\sim\Omega^{1/2}$ dependence.}
\label{f.LSW-XXZ}
\end{center}
\end{figure*}

\section{Linear spin-wave predictions for the XXZ model}

In this section we show the predictions of LSW theory for the scaling of ${\rm Var}(J^z)$ and the elementary excitation gap $\Delta E/{\cal J}$ for the XXZ model on a broader spectrum of parameters than those considered in the main text, namely for a variable anisotropy $-1 < \Delta \leq 1$, as well as for variable range interactions
\begin{equation}
{\cal J}_{ij} = \frac{{\cal J}}{|{\bm r}_i - {\bm r}_j|^\alpha}~.
\end{equation}
The case $\alpha=\infty$ recovers the limite of nearest-neighbor interactions, to which all the results presented in the main text, as well as elsewhere in the SM, apply. 
Fig.~\ref{f.LSW-XXZ} shows the $\Omega$ dependence of ${\rm Var}(J^z)$ and of the gap $\Delta E/{\cal J}$ in the case of $\alpha = \infty$ and $\alpha=3$ (dipolar interactions) for spins defined on a 2$d$ square lattice. We see that the prediction of a $\Omega^{1/2}$ dependence of both quantities in the limit $\Omega\to 0$ is robust upon varying the anisotropy $\Delta$ as well as the range of the interactions. ${\rm Var}(J^z)$ is found to grow with decreasing $\Delta$ (it becomes field-independent and stuck at the shot-noise value of $N/4$ in the limit $\Delta = -1$); while the gap decreases with $\Delta$. Longer-range interactions such as dipolar ones ($\alpha=3$, relevant for Rydberg atoms, trapped ions, magnetic atoms and dipolar molecules) lead to a larger gap and smaller variances compared to the nearest-neighboring case. 
  
\begin{figure*}[ht!]
\begin{center}
\includegraphics[width=\textwidth]{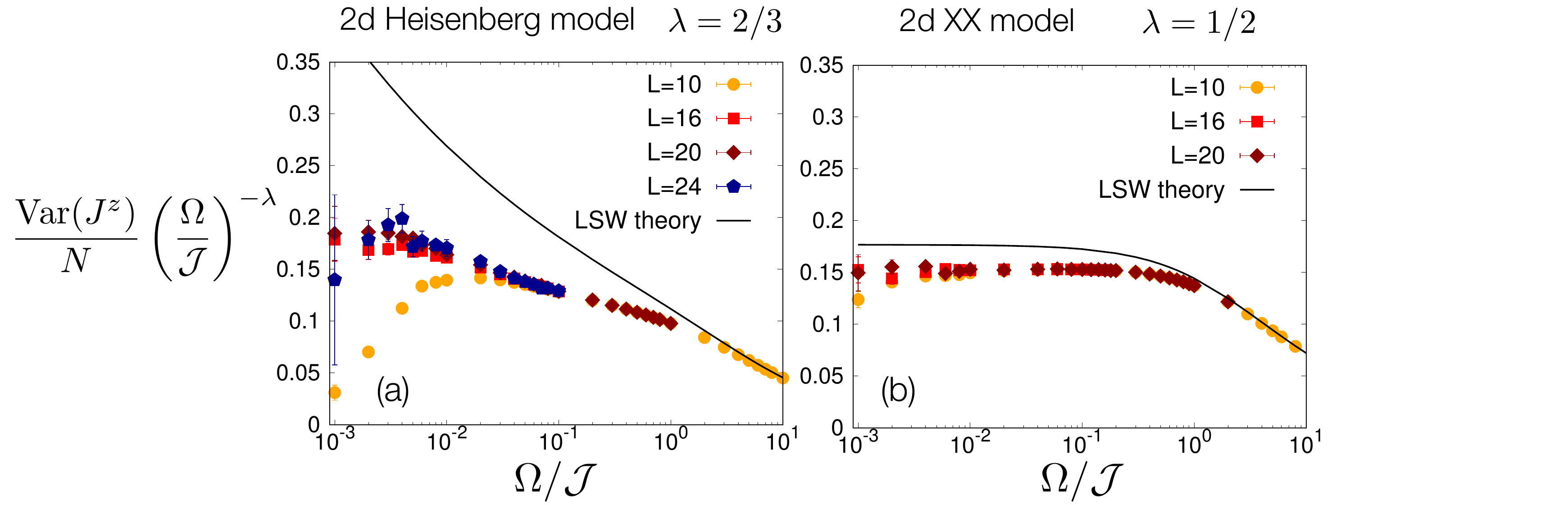}
\caption{Variance of the squeezed collective-spin component ${\rm Var}(J^z)$ as a function of the applied field for (a) the 2$d$ Heisenberg model and (b) the 2$d$ XX model. 
The symbols are QMC data, and the solid line corresponds to  LSW theory. 
The variance is multiplied by the factor $(\Omega/{\cal J})^\lambda$ so as to single out its power-law vanishing behavior as $\Omega \to 0$. }
\label{f.VarJz2dXXX}
\end{center}
\end{figure*}

\begin{figure*}[ht!]
\begin{center}
\includegraphics[width=0.75\textwidth]{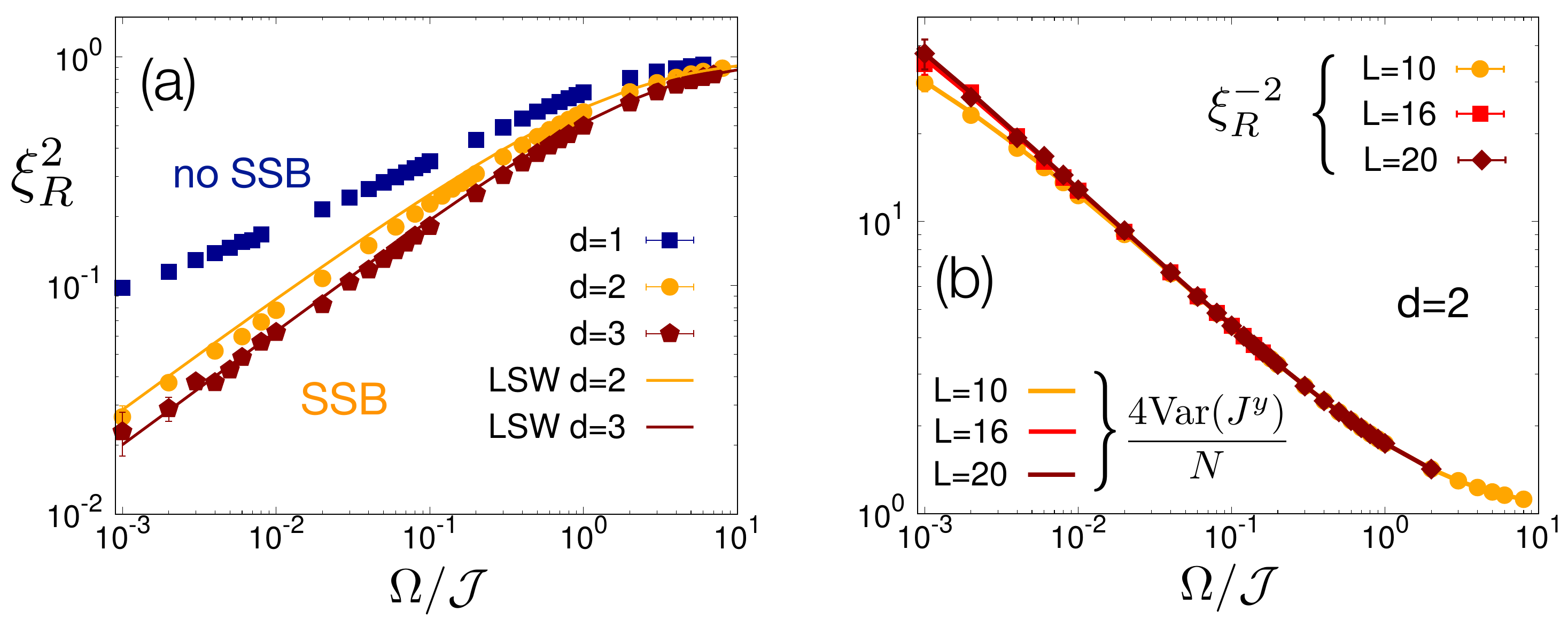}
\caption{Adiabatic squeezing in the $d$-dimensional XX model. (a) Spin squeezing parameter for various dimensions from QMC simulations: $\xi_R^2$ is found to scale to zero as ${\Omega}^{1/2}$ in the presence of spontaneous symmetry breaking (SSB) in the ground state for $d=2$ and 3, while it vanishes slowlier in its absence ($d=1$). Solid lines show the prediction of LSW theory. (b) Comparison between the inverse spin squeezing parameter $\xi_R^{-2}$ and $4 {\rm Var}(J^y)/N$ for the 2$d$ XX model for various system sizes. The coincidence of the two quantities shows that the ground state of the system is a minimal-uncertainty quantum state for all the field values we explored.}
\label{f.2dXX}
\end{center}
\end{figure*}

The predictions of LSW are fully corroborated by numerically exact results (see Figs.~\ref{f.gap}, \ref{f.VarJz2dXXX} and \ref{f.2dXX}(a), as well as Fig. 3(a) of the main text) -- with the exception of the special point $\Delta = 1$ in 2d). Hence we conclude that the scaling laws -- for the achievable spin squeezing and for the duration of an adiabatic preparation protocol -- quoted in the main text apply universally to the XXZ model whenever its ground state breaks a continuous U(1) (or SU(2)) symmetry and the order parameter does not commute with the Hamiltonian.

\section{Beyond-spin-wave behavior of spin squeezing in the 2d Heisenberg model}
\label{s.2dLSW}

Here we discuss in more details the field scaling of the variance of the squeezed spin component, ${\rm Var}(J^z)$, in the 2$d$ Heisenberg model  and in the 2$d$ XX model. 
Fig.~\ref{f.VarJz2dXXX} shows the field dependence of  ${\rm Var}(J^z)$ multiplied by a factor $(\Omega/{\cal J})^{-\lambda}$. A $\lambda$ exponent which leads to a field-independent product corresponds to the field-scaling exponent of 
${\rm Var}(J^z)$ at low fields. While LSW theory predicts $\lambda = 1/2$, we clearly observe that this exponent is inadequate for the low-field behavior of the 2$d$ Heisenberg model -- see Fig.~\ref{f.VarJz2dXXX}(a). Instead the heuristic choice $\lambda = 2/3$ seems to be more appropriate, and we retain this as the low-field scaling of ${\rm Var}(J^z)$, clearly faster than the LSW prediction. This picture is to be contrasted with that of the 2d XX model -- see Fig.~\ref{f.VarJz2dXXX}(b) -- for which the low-field scaling of ${\rm Var}(J^z)$ is captured very well by the LSW prediction (within a $\sim 10\%$ accuracy). 

\begin{figure}[ht!]
\begin{center}
\includegraphics[width=0.8\columnwidth]{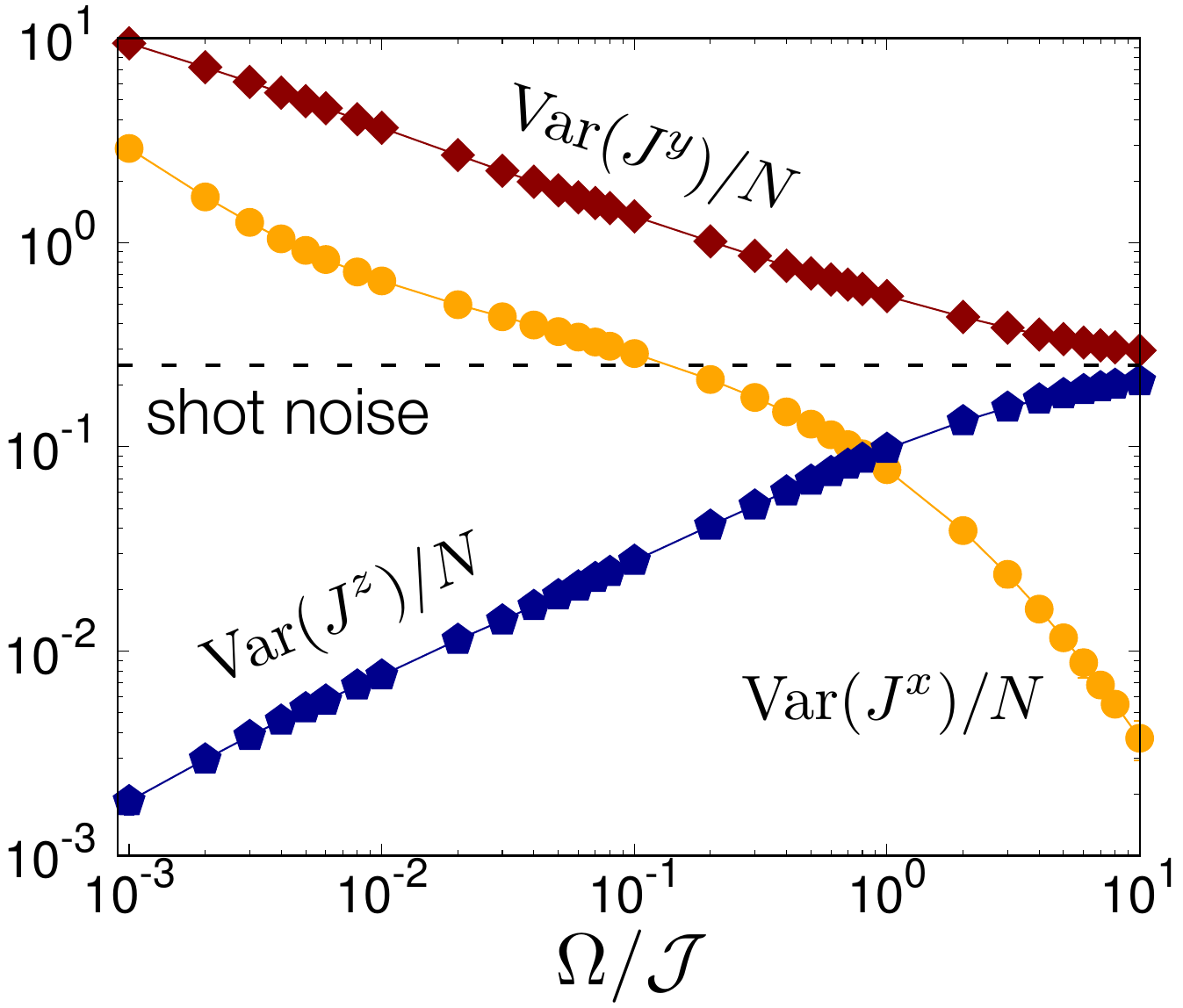}
\caption{Variance of the three components of the collective spin in the ground state of the 2$d$ Heisenberg model (QMC data for a $L=20$ lattice). The dashed line marks the shot-noise limit ${\rm Var}(J^\alpha)/N=1/4$.}
\label{f.variances}
\end{center}
\end{figure}

\section{QMC results for the XX model}

In this section we show our results for the XX model ($\Delta = 0$), relevant \emph{e.g.} for the physics of superconducting circuits \cite{Mi2021} as well of ultracold spinful atoms in optical lattices \cite{Jepsen2020}. Fig.~\ref{f.2dXX}(a) shows the field scaling of the spin squeezing parameter for the models in $d=1, 2$ and 3 dimensions. Similarly to the case of the Heisenberg model discussed in the main text, the scaling of the spin squeezing parameter as $\xi_R^2 \sim \Omega^{1/2}$ is observed in $d=2, 3$, {in agreement with the prediction of spin-wave theory. Interestingly, it is also observed in the case $d=1$, although it is significantly slower. This aspect is related to the possibility that certain Luttinger liquids (of which the XX chain is a realization) can develop scalable adiabatic spin squeezing. We will discuss this aspect in a future work.} 

Fig.~\ref{f.2dXX}(b) also shows that for $d=2$ the ground state at finite $\Omega$ is a state of minimal spin uncertainty, for which $\xi^{-2}_R \approx 4 {\rm Var}(J^y)/N$.  Similar result is also found in $d=3$.

\section{Ground-state variance of the collective spin components}

In Fig.~\ref{f.variances} we compare the variance of the three spin components $J^x$, $J^y$ and $J^z$ in the ground state of the 2$d$ Heisenberg model in a finite $\Omega$ field. We clearly observe that ${\rm Var}(J^y)$ is the largest among the three, meaning that the highest sensitivity of the state to rotations is achieved when the rotation axis is the $y$ axis. Also, among all operators of the form $O = \sum_i o_i$ (with $o_i$ a local operator associated to a finite neighborhood of site $i$) $J^y = \sum_i S_i^y$ is arguably the one that has the largest variance: this can be deduced from the fact that the variance is the integral of the correlation function ${\rm Var}(O) = \sum_{ij} \langle o_i o_j \rangle - \langle o_i \rangle \langle o_j \rangle$, and the correlation function which has the slowest spatial decay, leading to the largest integral, is the one of the order parameter in the SSB mechanism, corresponding to any spin component in the xy plane in the limit $\Omega \to 0$.

\begin{figure*}[ht!]
\begin{center}
\includegraphics[width=0.9\textwidth]{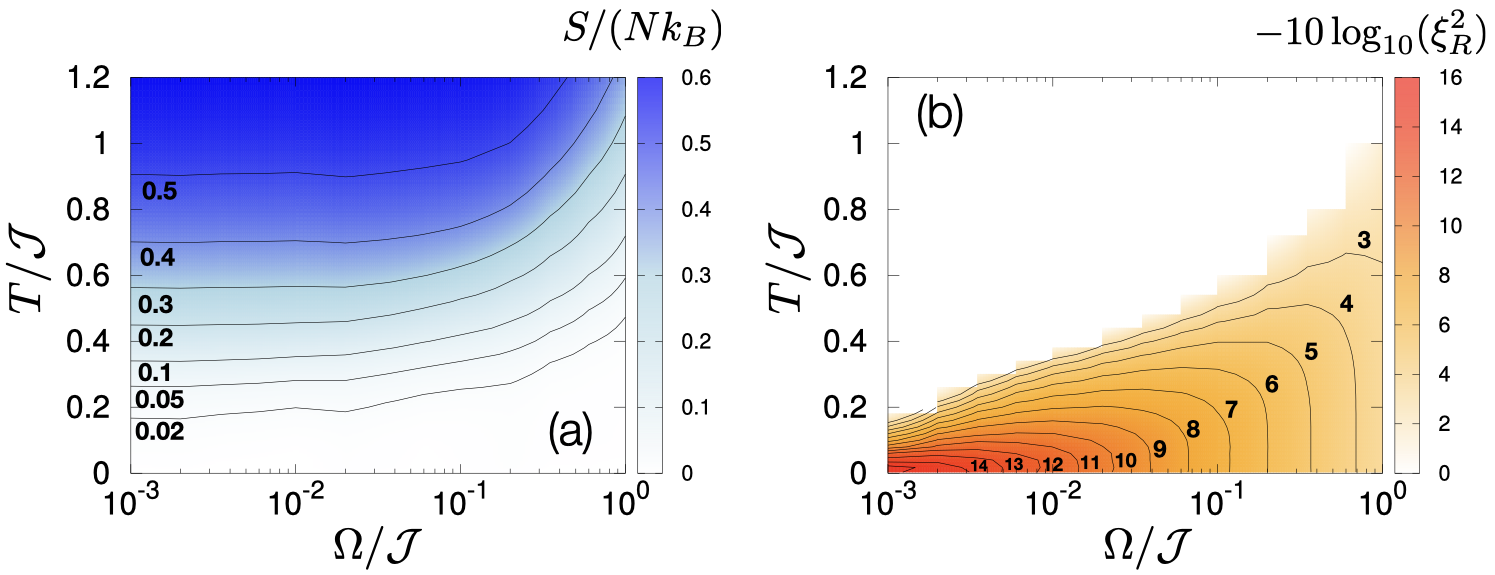}
\caption{Entropy and squeezing maps for the 2$d$ Heisenberg model. (a) Entropy map in the $(\Omega/{\cal J},T/{\cal J})$ plane; (b) squeezing map in the $(\Omega/{\cal J},T/{\cal J})$ plane.
All the QMC data shown here were obtained for a $L=24$ lattice.}
\label{f.entropy}
\end{center}
\end{figure*}

\section{Adiabatic squeezing at finite entropy}

Entropy curves can be obtained via QMC using a simple scheme of linear interpolation on the energy data. 
A direct result of the QMC calculation is the average energy per spin $e(T) = \langle \cal H \rangle/N$.  From a sufficiently fine grid $\{T_k\}$ of temperature values, we can then reconstruct the $e(T)$ curve by linear interpolation between two successive temperatures $e(T_k)$ and $e(T_{k+1})$, and therefore obtain the specific heat at the mid-point temperature $T_{k+1/2} = (T_{k+1} +T_k)/2$ as
\begin{equation}
c(T_{k+1/2}) \approx \frac{e(T_{k+1}) - e(T_k)}{T_{k+1} - T_k} ~.
\end{equation}
In order to obtain the $s(T) = S(T)/(Nk_B)$ curve via integration of the specific-heat data, we can proceed by a linear interpolation of the specific heat between midpoints
\begin{equation}
c(T) \approx c(T_{k+1/2}) + \frac{c(T_{k+3/2}) - c(T_{k+1/2})}{T_{k+3/2} - T_{k+1/2}} (T - T_{k+1/2})  
\end{equation}
valid for  $T_{k+1/2} \leq T \leq T_{k+3/2}$. This then allows us to estimate the entropy increment between two temperatures of the grid $\{T_k\}$ as 
\begin{equation}
\Delta s_k = s(T_{k+1}) - s(T_k) = \delta s_{k,<} + \delta s_{k,>}
\end{equation}
where 
\begin{equation}
\begin{split}
& \delta s_{k,<}  =  \int_{T_k}^{T_{k+1/2}} dT~ \frac{c(T)}{T} \\
& \approx  
 \left [ c(T_{k-1/2}) - \frac{c(T_{k+1/2}) - c(T_{k-1/2})}{T_{k+1/2} - T_{k-1/2}} \right ] \log\left(\frac{T_{k+1/2}}{T_{k}}\right)   \\
& +  \frac{c(T_{k+1/2}) - c(T_{k-1/2})}{T_{k+1/2} - T_{k-1/2}} (T_{k+1/2} - T_k)
\end{split}
\end{equation}
and 
\begin{equation}
\begin{split}
&\delta s_{k,>}  =  \int_{T_{k+1/2}}^{T_{k+1}} dT ~ \frac{c(T)}{T} \\
 &\approx  
\left [ c(T_{k+1/2}) - \frac{c(T_{k+3/2}) - c(T_{k+1/2})}{T_{k+3/2} - T_{k+1/2}} \right ] \log\left(\frac{T_{k+1}}{T_{k+1/2}}\right)   \\
& +  \frac{c(T_{k+3/2}) - c(T_{k+1/2})}{T_{k+3/2} - T_{k+1/2}} (T_{k+1} - T_{k+1/2})~.
\end{split}
\end{equation}
 
We have verified that this entropy reconstruction scheme delivers a similar result as that of a high-order polynomial fit of the $e(T)$ curve, as used \emph{e.g.} in Ref.~\cite{Carcyetal2021}; yet it has the advantage that no fitting procedure is involved. 

Fig.~\ref{f.entropy}(a) shows the resulting entropy map for the 2$d$ Heisenberg model as a function of temperature and applied field. We then combine the entropy map with the squeezing map at finite temperatures shown in Fig.~\ref{f.entropy}(b), to reconstruct the squeezing dependence on field and entropy, reported in Fig.~3 of the main text.


\begin{figure}[ht!]
\begin{center}
\includegraphics[width=0.85\columnwidth]{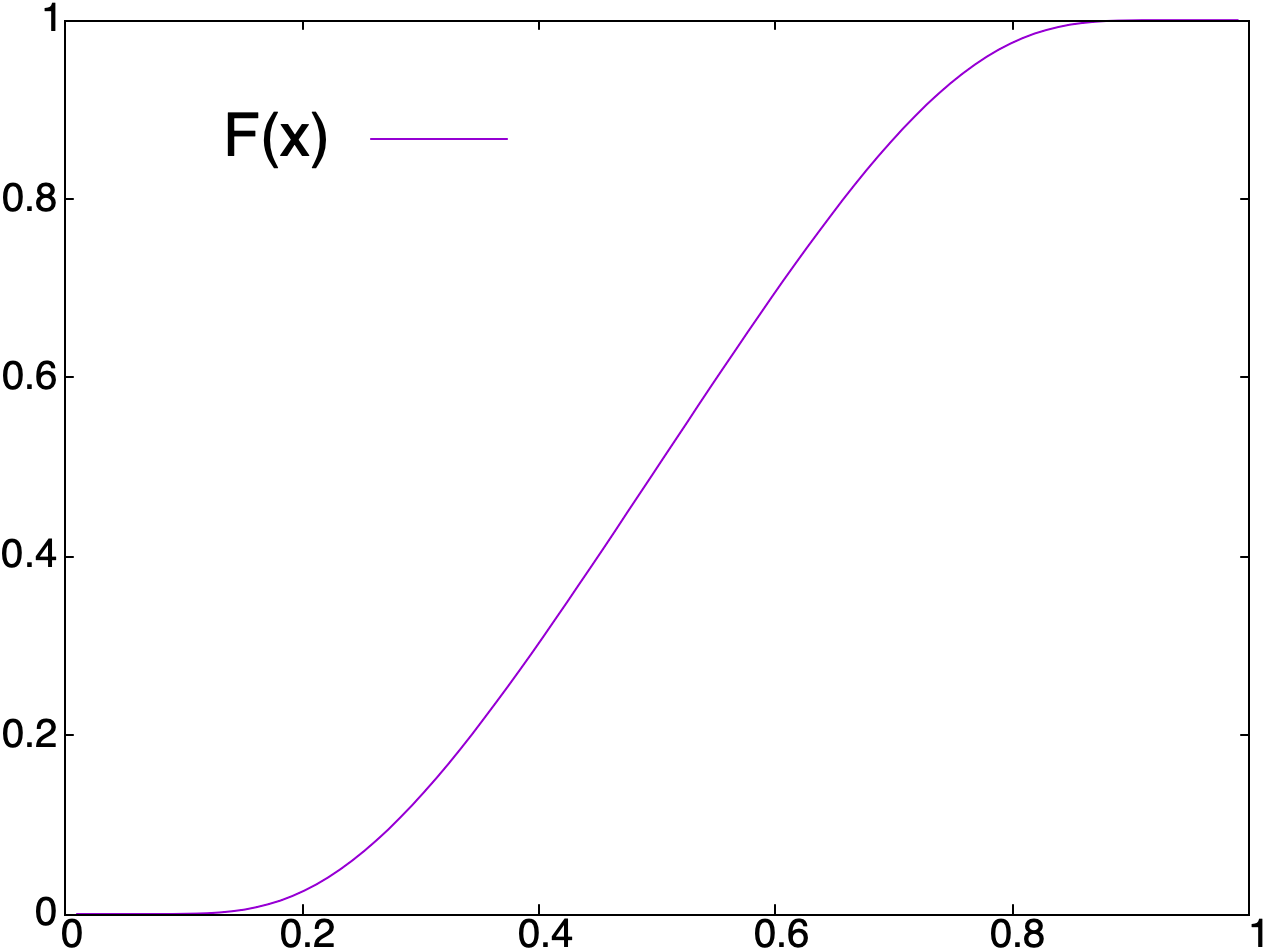}
\caption{Function $F(x)$ used for the quasi-adiabatic ramps.}
\label{f.Fx}
\end{center}
\end{figure}

\begin{figure*}[ht!]
\begin{center}
\includegraphics[width=0.9\textwidth]{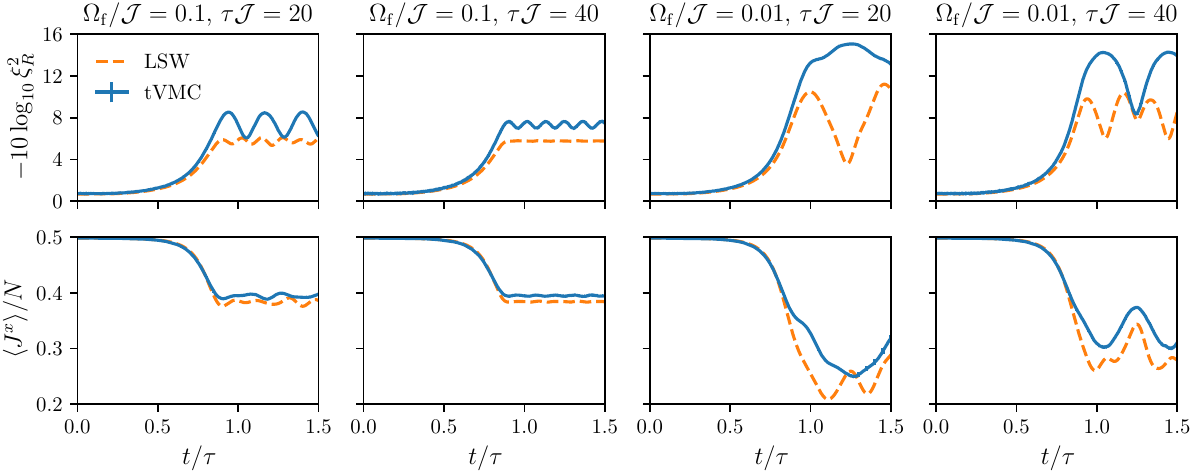}
\caption{Comparison between tVMC and LSW results for the evolution of the squeezing parameter $\xi_R^2$ (upper row) and of the magnetization per spin $\langle J^x \rangle/N$ (lower row) along quasi-adiabatic ramps starting from the ground state at $\Omega_{\rm i} = 10 {\cal J}$ and ending at two different values $\Omega_{\rm f} = 0.1 {\cal J}$ and $0.01 {\cal J}$, with two different ramp durations, $\tau {\cal J} = 20$ and 40. }
\label{f.tLSWtVMC}
\end{center}
\end{figure*}

\section{Time-dependent spin-wave theory vs. time-dependent variational Monte Carlo}

In order to corroborate the results coming from time-dependent variational Monte Carlo (tVMC), shown in the main text, we have compared them with an independent calculation, based on time-dependent LSW theory. 
This comparison is interesting especially at high $\Omega$ fields, for which, as we have seen in the main text and in the results above, the predictions of (static) LSW theory are fully quantitative. 

As in the main text, we consider an evolution which starts from the ground state at an initial field $\Omega_{\rm i} = 10 {\cal J}$, and ends at a final value $\Omega_{\rm f}$ with schedule $\Omega(t) = \Omega_{\rm i} + F(t/\tau) (\Omega_{\rm f}-\Omega_{\rm i})$, where $F(x)$ is reported in the main text, and shown in Fig.~\ref{f.Fx}. The time dependence of the field $\Omega(t)$ translates into a time-dependent $A_{\bm k}(t)$ parameter entering in the linear bosonic Hamiltonian 
Eq.~\eqref{e.LSW}. The evolution of the Gaussian state of the bosonic variables within LSW is fully described by the two correlation functions $G_{\bm k} = \langle b_{\bm k}^\dagger b_{\bm k} \rangle$ and $F_{\bm k} = \langle b_{\bm k} b_{-\bm k} \rangle$, which evolve according to the equations (descending from the Heisenberg equations for the $b_{\bm k}$ operators): 
\begin{eqnarray}
\frac{dG_{\bm k}}{dt} & = & -2 ~(1+\delta_{\bm k,0}) ~B_{\bm k} ~{\rm Im}(F_{\bm k})  \\
\frac{dF_{\bm k}}{dt} & = & -i (1+\delta_{\bm k,0}) \left [ 2 A_{\bm k}(t) F_{\bm k} + B_{\bm k} (1 + G_{\bm k} + G_{-\bm k} ) \right ] \nonumber
\end{eqnarray}
and from which our main observables of interest can be deduced
\begin{eqnarray}
\langle J^x \rangle & = & \frac{N}{2} - \sum_{\bm k} G_{\bm k} \nonumber \\
{\rm Var}(J^z) & = & \frac{1}{4} \left [ 1 + G_{\bm k} + G_{-\bm k} + 2 {\rm Re}(F_{\bm k}) \right ]  ~.
 \end{eqnarray}
 
Fig.~\ref{f.tLSWtVMC}(a)-(b) shows the comparison between the results of tVMC and those of LSW theory, obtained for the 2$d$ Heisenberg model with $L=12$, for final fields $\Omega_f = 0.1 {\cal J}$ and $0.01 {\cal J}$ and ramp durations $\tau {\cal J} = 20, 40$  (same as those shown in Fig. 3 in the main text). In particular we show the evolution of the spin squeezing parameter $\xi_R^2$ and of the magnetization per spin $\langle J^x \rangle/N$. We observe that tVMC and LSW agree perfectly for most of the ramp duration, while they deviate upon approaching the end of the ramp ($t = \tau$) and in the subsequent time evolution, during which $\Omega$ is held fixed at its final value. This deviation is easily understood within an adiabatic picture, namely from the fact that (static) LSW theory overestimates the $\xi_R^2$ parameter (as shown in the main text), namely it underestimates $-10 \log_{10} (\xi_R^2)$ shown in the figure; while the tVMC results for quasi-adiabatic ramps produce values of the $\xi_R^2$ which are systematically closer to the (static) QMC estimate. From this comparison we conclude therefore that our tVMC results, while not numerically exact, are fully quantitative, as they reproduce the LSW results at large fields ($t < \tau$); and, for quasi-adiabatic ramps, they oscillate around a value compatible with the QMC prediction for $t > \tau$.

\section{Spin squeezing of a bosonic lattice clock: the case of $^{174}$Yb}

In this section we discuss how the collisional properties of $^{174}$Yb, along with its clock transition, would make of this bosonic atom an ideal candidate for the implementation of our squeezing protocol within a quantum metrology platform. As discussed in the main text, the choice of a bosonic atom allows for the protocol to be implemented in its simplest form, namely starting from a uniform coherent spin state and driven by a uniform Rabi field. Following Refs.~\cite{Bouganneetal2017} and \cite{Franchietal2017}, we consider the clock transition between the two states $|g\rangle = ~^1S_0$ and $|e\rangle = ~^3P_0$. The collisional properties of these two states have been fully characterized in the same references, with scattering lengths $a_{gg} \approx 105 ~a_0$, $a_{eg} \approx  86 ~a_0$ and $a_{ee} \approx 102 ~a_0$ in units of the Bohr radius $a_0$ from Ref.~\cite{Bouganneetal2017}; and  $a_{gg} \approx 105 ~a_0$, $a_{eg} \approx  95 ~a_0$ and $a_{ee} \approx 127 ~a_0$ from Ref.~\cite{Franchietal2017}. When atoms are loaded in a deep optical lattice at the magic wavelength of 759.4 nm -- at which the optical potential becomes independent of the internal state -- the two atomic states have the same hopping parameter $\lambda$, and the single-band interaction parameters $U_{\alpha\beta}$ (with $\alpha, \beta = e, g$) are simply proportional to the scattering lengths. The effective spin model realized when atoms are prepared in a Mott insulating state has parameters \cite{Duanetal2003}
\begin{equation}
{\cal J} = \frac{\lambda^2}{U_{eg}}  ~~~~~~ {\cal J}\Delta = \lambda^2\left (\frac{1}{U_{gg}} + \frac{1}{U_{ee}} - \frac{1}{U_{eg}} \right ) ~.
\end{equation}
 By inserting the scattering lengths determined by Ref.~\cite{Bouganneetal2017}, we find that $\Delta \approx - 0.66$; whereas when using the scattering lengths determined by Ref.~\cite{Franchietal2017} we find $\Delta \approx -0.65$. In both cases, the $\Delta$ parameter falls well within the regime of $- 1 < \Delta \leq 1$ of validity of the adiabatic squeezing protocol.
 
 This suggests that, based on the approach proposed in this work, Mott insulators of $^{174}$Yb may offer a most promising platform for the realization of scalable spin squeezing in a bosonic optical lattice clock.

\bibliography{refs.bib}

\end{document}